# Imaging an isolated water molecule using a single electron wave packet

Xinyao Liu[1], Kasra Amini[1,2,†], Tobias Steinle[1], Aurelien Sanchez[1], Moniruzzaman Shaikh[1], Blanca Belsa[1], Johannes Steinmetzer[3], Anh-Thu Le[4,5], Robert Moshammer[6], Thomas Pfeifer[6], Joachim Ullrich[6,7], Robert Moszynski[2], C.D. Lin[4], Stefanie Gräfe[3,8], Jens Biegert[1,9]

[1]ICFO - Institut de Ciencies Fotoniques, The Barcelona Institute of Science and Technology, 08860 Castelldefels (Barcelona), Spain.
[2]Department of Chemistry, University of Warsaw, 02-093 Warsaw, Poland.
[3]Institute of Physical Chemistry, Friedrich-Schiller University, 07743 Jena, Germany.
[4]Department of Physics, J. R. Macdonald Laboratory, Kansas State University, 66506-2604 Manhattan, KS, USA.
[5]Missouri University of Science and Technology, Rolla, MO 65409, USA.
[6]Max-Planck-Institut für Kernphysik, Saupfercheckweg 1, 69117, Heidelberg, Germany.
[7]Physikalisch-Technische Bundesanstalt (PTB), D-38116 Braunschweig, Germany.
[8]Abbe Center of Photonics, Friedrich-Schiller University, 07745 Jena, Germany.
[9]ICREA, Pg. Lluís Companys 23, 08010 Barcelona, Spain.
[†]To whom correspondence should be addressed to. Email: kasra.amini@icfo.eu.

**Abstract**
Observing changes in molecular structure requires atomic-scale Ångstrom and femtosecond spatio-temporal resolution. We use the Fourier transform (FT) variant of laser-induced electron diffraction (LIED), FT-LIED, to directly retrieve the molecular structure of $H_2O^+$ with picometre and femtosecond resolution without *a priori* knowledge of the molecular structure nor the use of retrieval algorithms or *ab initio* calculations. We identify a symmetrically stretched $H_2O^+$ field-dressed structure that is most likely in the ground electronic state. We subsequently study the nuclear response of an isolated water molecule to an external laser field at four different field strengths. We show that upon increasing the laser field strength from 2.5 to 3.8 V/Å, the O-H bond is further stretched and the molecule slightly bends. The observed ultrafast structural changes lead to an increase in the dipole moment of water and, in turn, a stronger dipole interaction between the nuclear framework of the molecule and the intense laser field. Our results provide important insights into the coupling of the nuclear framework to a laser field as the molecular geometry of $H_2O^+$ is altered in the presence of an external field.

## I - Introduction

Water plays a crucial role in many physical, chemical and biological processes.[1] In fact, modifying the geometric structure of water can, for example, influence the folding dynamics of proteins surrounded by water solvation shells.[2-4] Such a modification of molecular structure can be accomplished by exposing molecules to strong fields with field strengths comparable to the Coulombic attraction between electrons and protons. Field-dressing the molecule can lead to a perturbation in its potential energy surfaces, and in turn could lead to a new energy minimum and possible modification of the equilibrium molecular structure on the nuclear (*i.e.* femtosecond; 1 fs = $10^{-15}$ s) timescale. In fact, external fields with field strengths of around 0.5 V/Å have been theoretically reported to already cause structural changes in a water molecule.[5] Moreover, it is also reported that each water molecule, in the absence of an external field, experiences a local electric field of 2 V/Å from nearby water monomers in liquid water.[6] Exactly how structural deformation of water molecules arises was found to strongly depend on the orientation of the molecule's dipole moment relative to the electric field vector of the external field with field strengths of between 0.5 - 2.5 V/Å:[5] if the dipole moment of the molecule is aligned with (against) the external electric field, then the molecule is stretched and bent (contracted and straightened) with increasing field strength.[5] Although experimental studies into the structural retrieval of $H_2O$ and its isotopologue $D_2O$ have been reported with femtosecond imaging techniques such as Coulomb explosion imaging, very intense femtosecond laser pulses (>5 x $10^{15}$ $Wcm^{-2}$; >19 V/Å) were used which led to significant structural deformation.[7,8] Such ultraintense femtosecond laser pulses in fact generate highly-charged molecular water cations that undergo extremely fast nuclear motion due to the significant Coulomb repulsion that is present in the multiply-charged molecular cation (*i.e.* "Coloumb explosion"). Sanderson *et al.* (1999) reported that exposing gas-phase water molecules to an ultraintense (>$10^{16}$ $Wcm^{-2}$), 50 fs (FWHM) 0.79 μm laser pulse led to the detection of significantly straightened ($\Phi_{HOH}$ = 130-180°) multiply charged $H_2O^{3+}$ and $H_2O^{4+}$ ions. Légaré *et al.* (2005) retrieved a $D_2O^{4+}$ structure that is symmetrically stretched ($R_{OD}$ = 1.24 Å) and straightened ($\Phi_{DOD}$ = 117°) in the presence of an ultraintense (5 x $10^{15}$ $Wcm^{-2}$), 8 fs (FWHM) 0.8 μm laser pulse. Structural information was, in the case discussed so far, only indirectly obtained. In this contribution, we obtain direct structural information utilizing laser-induced electron diffraction (LIED).[8-Error! Reference source not found.] In LIED, an electron wave packet (EWP) is: (i) ejected from the target system (*e.g.* an isolated $H_2O$ molecule) by strong-field tunnel ionization; (ii) accelerated and returned by the oscillating laser field; and (iii) scattered against the atomic cores of the target ion. Structural information is embedded onto the momentum distribution of the detected scattered electrons. Thus, using the target's own EWP as a structural probe permits a snapshot of the isolated target's structure to be captured with femtosecond and sub-Ångstrom spatiotemporal resolution. Analogous non-laser-induced electron diffraction methods to LIED, such as ultrafast electron diffraction[20,21], can also capture structural information with Ångstrom spatial resolution[22,23], although they are typically limited to a temporal resolution of several hundreds of femtoseconds.

Here, we investigate the nuclear response of an isolated neutral water molecule exposed to an external laser field of field strengths that are typically experienced in liquid water under natural conditions. Specifically, we use intense (>$10^{13}$ $Wcm^{-2}$), 97 fs (FWHM) 3.2 μm laser pulses with field strengths of 2.5 – 3.8 V/Å to study field-dependent changes in the molecular structure which are spatially resolved on the atomic (*i.e.* Ångstrom; 1 Å = $10^{-10}$ m) scale. We use LIED to directly retrieve structural information of an isolated $H_2O^+$ cation with picometre (pm; 1 pm = $10^{-12}$ m) and femtosecond spatio-temporal resolution. We directly retrieve a symmetrically stretched and slightly straightened $H_2O^+$ structure which is indicative of a $H_2O^+$ structure in the ground electronic state. Our measurements also reveal that increasing the laser field strength from 2.5 V/Å to 3.8 V/Å in fact leads to an observed increase in the O-H (H-H) internuclear distance of 14% to 35% (17% to 35%) relative to the equilibrium field-free $H_2O^+$ ground state structure.

This paper is organized as follows: a brief overview of the experimental setup and the theoretical methods employed in this work is given in Section II, with the results presented in Section III followed by a summary of the experimental and theoretical results in Section IV.

## II – Methodologies

### IIa – Experimental

Our experimental set-up has been previously described in detail elsewhere with only a brief summary provided here.[24] Helium gas (~300 mbar) was passed through a temperature-controlled reservoir setup containing pure deionized liquid water (~50 mL; Sigma Aldrich). The vaporized gas-phase water molecules picked up by the helium carrier gas were supersonically expanded into the jet chamber and passed through two skimmer stages. The collimated molecular beam of gas-phase water molecules was then intercepted with an orthogonally directed 3.2 μm linearly polarized laser pulse. The 3.2 μm laser pulse was obtained (pulse duration of 97 fs (FWHM), and a pulse energy of 8.4 – 15.6 μJ) from a home-built optical parametric chirped pulse amplifier (OPCPA) set-up with up to 21 W output power at a repetition rate of 160 kHz.[25,26] The laser pulse was focused into the molecular beam using an on-axis paraboloid that is placed inside of the reaction microscope, and is focussed down to a focal spot of 6 – 7 μm. In this paper, we performed measurements at four different peak pulse intensities of between $8.5 \times 10^{13} - 1.9 \times 10^{14}$ Wcm$^2$. We operated under the quasistatic ionization regime as given by the Keldysh parameter of $\gamma \sim 0.2$.[27] Rescattered electrons were generated in four separate FT-LIED measurements with a ponderomotive energy ($U_\text{p} \propto I_0\lambda^2$; *i.e.* the average kinetic energy of a free electron in an oscillating electric field) of between 81 – 185 eV, which corresponded to the maximum classical returning energy $E_\text{r}^{\max}$ (= 3.17$U_\text{p}$) of between 257 – 586 eV, and the maximum backscattered energy $E_\text{resc}^{\max}$ (= 10$U_\text{p}$) of 810 – 1,850 eV. Using homogenous electric and magnetic fields, the resulting electrons and ions were extracted out of the interaction region and projected towards two separate delay-line anode detectors. The full three-dimensional momentum distribution of electrons and ions was simultaneously detected in coincidence with full 4π detection using a reaction microscope (ReMi), enabling the isolation of the reaction path of interest.[28]

### IIb - Theoretical

The field-free potential energy surface (PES) for the $^2B_1$ electronic ground state of $H_2O^+$ was obtained with the complete active space self consistent field method (CASSCF) and the aug-cc-PVTZ basis set.[29-31] All calculations were completed using the DALTON2018.0 program suite.[32] A $C_{2v}$ symmetry was employed for all calculations. The molecule was oriented to lie in the *yz*-plane with the *z*-axis coinciding with the $C_2$ rotation-axis. An active space of 7 electrons in 8 orbitals was selected for $H_2O^+$. A (4,2,2,0) partitioning of the active space was used with the four numbers corresponding to the active orbitals in the respective representations of the $C_{2v}$ group ($A_1$, $B_1$, $B_2$, $A_2$). The PES is spanned by two internal coordinates: (i) the H-O-H angle was varied from 180° to 55° in steps of 1°; and (ii) the O-H bond length was varied from 0.60 Å to 2.25 Å in steps of 0.05 Å. A symmetric O-H stretch was assumed. Permanent dipole moments ($\mu_x$, $\mu_y$, $\mu_z$) were calculated for all points of the PES using the response theory approach as implemented in DALTON.[33] Results for selected geometries obtained at different field strengths between 2.5 – 3.8 V/Å are given in Table 2.

### III - Results

**FT-LIED analysis.** The structural retrieval procedure is described here using exemplary LIED data recorded at a field strength of 3.4 V/Å ($U_p$ = 150 eV, $I_0$ = 1.6 x $10^{14}$ Wcm$^{-2}$). We employ the Fourier transform (FT) variant of LIED, called FT-LIED,[15] to directly retrieve structural information without *a priori* knowledge of the molecular structure nor the use of retrieval algorithms or *ab initio* calculations. In FT-LIED, similar to the previously reported fixed-angle broadband laser-driven electron scattering (FABLES),[14,Error! Reference source not found.] only the backrescattered highly-energetic LIED electrons are considered since the Fourier transform of the molecular interference signal embedded within their momentum distribution in the far-field directly provides an image of the molecular structure.

Fig. 1 shows the two-dimensional momentum distribution of longitudinal ($P_l$; parallel to the laser polarization) and transverse ($P_t$; perpendicular to the laser polarization) momenta for all electrons. We detect "direct" electrons, which escaped the laser field without recollision, and "rescattered" electrons, which were accelerated and returned by the laser field with recollision. For direct (rescattered) electrons, the detected maximal classical rescattering energy is $2U_p$ ($10U_p$), with the direct (rescattered) electrons dominating below (above) $2U_p$. To operate in the rescattering frame, the vector potential kick, $A_r$, that rescattered electrons experience in the presence of the strong mid-infrared laser field, is subtracted from the detected rescattered momentum, $k_{resc}$, to obtain the returning momentum, $k_r$, at the instance of rescattering. Since we implement the FT-LIED method, we only analyse electrons with $k_r \geq 2.4$ a.u. (*i.e.* $P_l \geq 4.7$ a.u.) and a rescattering angle, $\theta_r$, of 170 – 190° (*i.e.* rescattering cone of $\Delta\theta_r$ = 10° around the back-rescattering angle of $\theta_r$ = 180°). Specifically, we extract an energy-dependent interference signal by integrating an area indicated by a block arc in momentum space, as shown in Fig. 1, at various different vector potential kicks (see white, yellow and green arrows along $P_t$ = 0 in Fig. 1).

**Electron-ion 3D coincidence detection.** The interference signal extracted from Fig. 1 in fact corresponds to electrons generated not only from our ion of interest, $H_2O^+$, but also other, background ions. To avoid contamination of our interference signal from these background ions belonging to other competing processes, such as multi-photon ionization and Coulomb explosion processes, we implement electron-ion coincidence detection to isolate the LIED interference signal from only the $H_2O^+$ ion of interest. Fig. 2a shows the ion time-of-flight (ToF) spectrum of all the positively charged fragments detected in coincidence with their corresponding electron(s), with the dominant ToF peak at around 5.5 μs belonging to the molecular ion, $H_2O^+$.

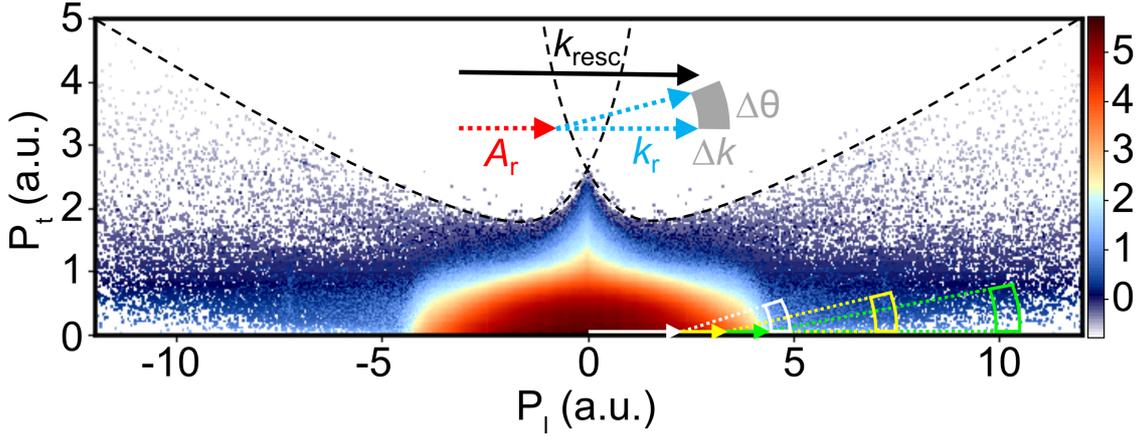

**Fig. 1 | FT-LIED extraction.** Two-dimensional map of the longitudinal, $P_l$, and transverse, $P_t$, momenta given in atomic units (a.u.). The return momentum, $k_r$, at the instance of rescattering is obtained by subtracting the vector potential, $A_r$, from the detected rescattering momentum, $k_{resc}$, given by $k_r = k_{resc} - A_r$. The energy-dependent interference signal is extracted at various different vector potential kicks (see white, yellow and green arrows along $P_t = 0$) by integrating the area indicated by a block arc. A schematic is drawn at the top of the figure relating $k_{resc}$ (black arrow), $k_r$ (blue arrows), $A_r$ (red arrow), and the integrated area indicated by the grey shaded block arc corresponding to a small range of rescattering angles and momenta, $\Delta\theta$ and $\Delta k$, respectively. We use $\Delta\theta_r$ and $\Delta k_r$ values of 10° and 0.2 a.u., respectively. The dotted black lines represent the detection acceptance angle of the reaction microscope.[28]

The ToF range corresponding to $H_2O^+$ ions is indicated by the blue shaded region shown in the inset of Fig. 2a. The total electron signal for (i) all electrons (blue dotted trace) and (ii) those electrons detected in coincidence with $H_2O^+$ (black solid trace) is shown in Fig. 2b. In both distributions, the $2U_p$ and $10U_p$ classical cut-offs are clearly visible (green arrows). Moreover, the inset in panel (b) highlights the advantage of coincidence detection, with the modulated interference signal appearing more pronounced in the $H_2O^+$ coincidence distribution than that of all electrons.

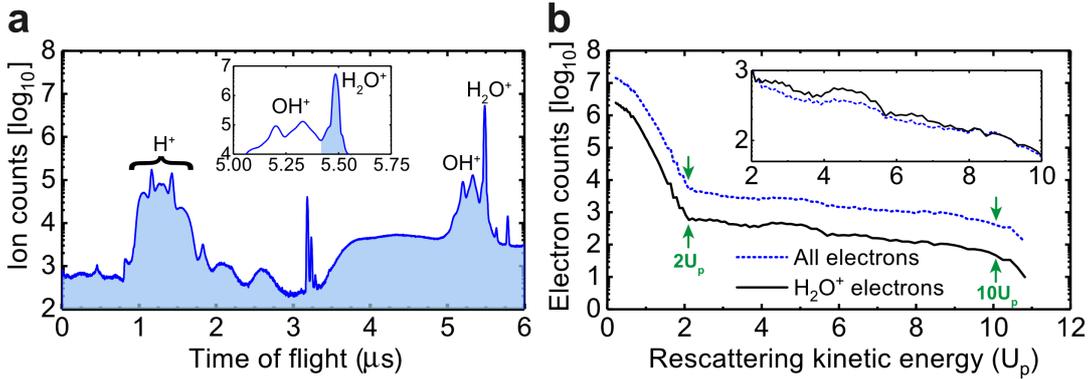

**Fig. 2 | Coincidence detection.** (a) The ion time-of-flight (ToF) spectrum, with the main ToF peak corresponding to the molecular ion, $H_2O^+$. The inset shows a zoomed-in view of the ToF spectrum at around 5.50 μs, with the blue shading indicating the ToF range of $H_2O^+$ ions. **(b)** Electron counts as a function of rescattered kinetic energy given in ponderomotive energy, $U_p$, for all electrons (blue dotted trace) and those electrons detected in coincidence with $H_2O^+$ (black solid trace). Here, the $2U_p$ and $10U_p$ classical cut-offs are indicated (green arrows) and correspond to a $U_p$ of 150 eV. The inset shows a close-up view of the electron signal between 2 – $10U_p$, with the two distributions overlaid on top of each other to highlight the advantage of coincidence imaging with the more pronounced modulated interference signal seen in the $H_2O^+$ case.

**Molecular structure retrieval.** Considering electrons detected in coincidence with $H_2O^+$, we generate a plot of electron counts as a function of return kinetic energy in the range of 80 - 460 eV corresponding to the rescattering plateau of $2U_p - 10U_p$ range, as shown in Fig. 3a. Such a plot is generated by integrating the grey block arc area as shown in Fig. 1 at different rescattering momenta, $k_{resc}$. We directly retrieve structural information from the modulated $H_2O^+$ electron signal in the rescattering regime. Here, the measured total interference signal ($I_T$; blue solid trace) from the backrescattered LIED electrons is plotted, which in fact contains contributions from both the coherent molecular interference signal ($I_M$) and the incoherent atomic signal ($I_A$). Here, $I_A$ is the incoherent sum of scattering the atoms, whilst $I_M$ is the coherent interference of scattering from the atoms. In fact, $I_A$ is independent of molecular structure, and can be empirically extracted by fitting a third-order polynomial function (black dotted trace) to the total interference signal. Subtracting $I_A$ from $I_T$ gives our interference signal of interest, $I_M$, which is sensitive to the molecular structure. To highlight changes in $I_T$ as a result of $I_M$, the molecular contrast factor (MCF) is calculated as given by

$$MCF = \frac{I_T - I_A}{I_A} = \frac{I_M}{I_A},$$

and is plotted in Fig. 3b as a function of momentum transfer, $q = 2 \cdot k_r$. The observed oscillations in the MCF (blue solid trace) provide a unique, sensitive signature of the molecular structure, with the shaded regions representing the estimated statistical error of the measurement. Fast Fourier transform (FFT) of the molecular interference signal embedded within the MCF generates an FFT spectrum, as shown in Fig. 3c. Before transforming, a Kaiser window and zero padding are applied.[33,36] In panel (c), the FFT spectrum (blue solid trace), individual Gaussian fits (black dotted traces) and the sum of the two Gaussian fits (black solid trace) are presented. The centre position of the individual Gaussian fits to the two FFT peaks appear at 1.24 ± 0.08 Å and 2.04 ± 0.08 Å, with the features above 3 Å arising from noise. Table 1 shows the O-H and H-H internuclear distances reported for field-free $H_2O$ in its ground electronic state and field-free $H_2O^+$ in the ground and first two excited electronic states. Comparing our FFT spectrum shown in Fig. 3c to the data in Table 1, it is clear that the first FFT peak at 1.24 ± 0.08 Å corresponds to the O-H internuclear distance ($H_2O(\widetilde{X})$: 0.96 Å; $H_2O^+(\widetilde{X})$: 1.00 Å), whilst the second FFT peak measured at 2.04 ± 0.08 Å corresponds to a significantly stretched H-H internuclear distance ($H_2O(\widetilde{X})$: 1.52 Å; $H_2O^+(\widetilde{X})$: 1.63 Å).[37-39]

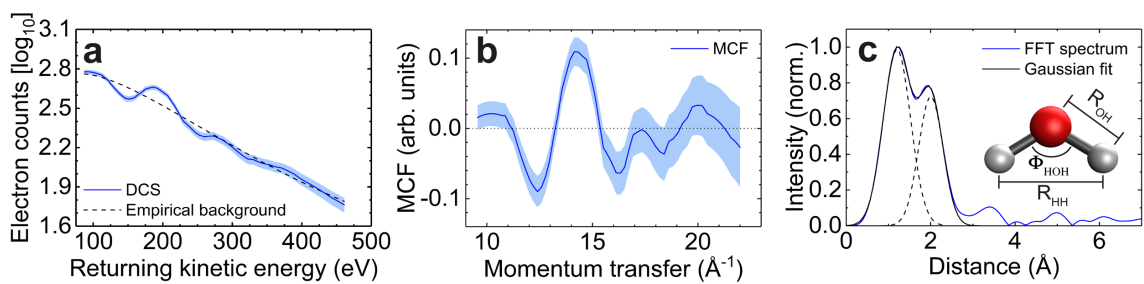

**Fig. 3 | Structure retrieval. (a)** The modulated total interference signal (blue trace) is shown with the background atomic signal (black dotted trace), the latter of which is empirically extracted using a third-order polynomial fit. The blue shaded region indicates the errors estimated via Poissonian statistics. **(b)** The molecular contrast factor (MCF) plotted as a function of momentum transfer. **(c)** Fast Fourier spectrum (blue trace) with Gaussian fits (black traces). A sum of two Gaussians (black solid trace) along with the individual Gaussian fits (black dashed traces) are presented. The following geometrical parameters are extracted: O-H internuclear distance, $R_{OH}$ = 1.24 ± 0.08 Å; H-H internuclear distance, $R_{HH}$ = 2.04 ± 0.08 Å; and H-O-H bond angle, $\Phi_{HOH}$ = 111 ± 7°. A sketch of the measured $H_2O^+$ structure is shown.

The FT-LIED measured internuclear distances give an H-O-H bond angle of $\Phi_{HOH}$ = 111 ± 7° which is in closer agreement with the ground state structure of $H_2O^+$ ($\Phi_{HOH}$ = 109°) than with neutral $H_2O$ ($\Phi_{HOH}$ = 105°).[37-39] The excited electronic states of neutral $H_2O$ are dissociative, and thus on the 7 – 9 fs timescale of the returning LIED electron, the molecular system will have stretched significantly (*i.e.* the $R_{OH}$ would be significantly greater than the 1-2 Å distance that we measure).[40,41] The second and third electronic excited state of the $H_2O^+$ cation are also dissociative.[42] Moreover, the equilibrium geometric structure of $H_2O^+$ in its first excited electronic state, particularly the $\Phi_{HOH}$ angle, is significantly different to the $H_2O^+$ ground state equilibrium structure, as shown in Table 1. The ionization potential ($I_p$) of the highest occupied molecular orbital (HOMO; $I_p$ = 12.6 eV) in neutral $H_2O$ is around 2 eV lower than that of the HOMO-1 ($I_p$ = 14.7 eV), and thus ionization from the HOMO would be expected to dominate.[43,44] On the other hand, several experimental and theoretical investigations have shown that the HOMO-1 in fact plays an appreciable role during the strong-field ionization of neutral $H_2O$, and should be considered (HOMO-2 contribution was found to be negligible).[44-46] Thus, it could be possible that our extracted FT-LIED $H_2O^+$ structure may arise from the removal of an electron from a superposition of the HOMO and HOMO-1 orbitals. Interestingly, Rao *et al.* (2015) not only theoretically further confirmed the same $D_2O^+/H_2O^+$ high-harmonic ratio trend observed experimentally by Farrell *et al.* (2011) with high-harmonic spectroscopy, Rao *et al.* (2015) also present the calculated time-resolved structures extracted from their time-dependent wave packet dynamical calculations propagating only on either the $\tilde{X}^2B_1$ ground electronic state or the $\tilde{A}^2A_1$ first excited state of $H_2O^+$. At around t = 7 – 8 fs (*i.e.* 7 – 8 fs after the ion was generated), Rao *et al.* (2015) report structures with a $\Phi_{HOH}$ of around 111° (168°) for $H_2O^+$ in the $\tilde{X}^2B_1$ ($\tilde{A}^2A_1$) electronic state. Since the return time of the LIED electron wave packet from the moment of ionization to the moment of rescattering is roughly 7 – 9 fs, the FT-LIED structure that we report ($\Phi_{HOH}$ = 111 ± 7°) seems to best agree with the $H_2O^+$ structure in the $\tilde{X}^2B_1$ ground electronic state at roughly the same timescale reported by Rao *et al.* (2015).

**Table 1 | Equilibrium geometrical parameters of $H_2O$ and $H_2O^+$.** The O-H and H-H internuclear distances, $R_{OH}$ and $R_{HH}$, respectively, and the H-O-H angle, $\Phi_{HOH}$, for field-free neutral $H_2O$ in the ground electronic state.[37,38] The same geometric parameters for field-free $H_2O^+$ cation in the ground and first excited electronic states are also presented.[39,42]

|  | $R_{OH}$ (Å) | $R_{HH}$ (Å) | $\Phi_{HOH}$ (°) |
|---|---|---|---|
| $H_2O$ ($\widetilde{X}^1A_1$) | 0.96 | 1.52 | 105 |
| $H_2O^+$ ($\widetilde{X}^2B_1$) | 1.00 | 1.63 | 109 |
| $H_2O^+$ ($\widetilde{A}^2A_1$) | 0.99 | 1.98 | 180 |

**Structure dependence on field strength.** We investigate the effect of field-dressing the molecule and the response of the nuclear framework to the strong laser field. Using the same analysis procedure as described above, the geometric parameters of $H_2O^+$ at three other laser field strengths of 2.5, 3.1, 3.8 V/Å are extracted, the results of which are presented in Fig. 4. The directly retrieved O-H (blue data) and H-H (red data) internuclear distances are shown in Fig. 4a with the corresponding H-O-H angle (black data) shown in Fig. 4b for four laser field strengths. The extracted geometrical parameters at 2.5, 3.1 and 3.8 V/Å further support our earlier assignment of the measured FT-LIED structure at 3.4 V/Å to $H_2O^+$ in the ground electronic state. Moreover, at these four laser field strengths, the O-H and H-H internuclear distances both increase with increasing field strength. In fact, increasing the laser field strength from 2.5 V/Å to 3.8 V/Å leads to an observed increase in the O-H (H-H) internuclear distance of 14% to 35% (17% to 35%) relative to the equilibrium field-free $H_2O^+$ ground state structure. A line of best fit using the least squares method is also applied to the respective data (dash-dot traces), where a linear increase in the O-H and H-H internuclear distances is accompanied by a slight decrease in the H-O-H angle as a function of increasing field strength. An illustration of the geometries directly retrieved at these four field strengths is presented in Fig. 4c.

**Mechanism of O-H bond length stretch.** The mechanism that leads to the distortion of the nuclear framework in field-dressed $H_2O$ is explained as follows. The optical period of our 3.2 μm laser (10.7 fs) is long enough to enable an appreciable dipole interaction of the molecule with the electric field corresponding to the laser field. Field-dressing the molecule leads to the lowering of the potential energy by μ*E in first-order approximation and a more energetically stable structure with a stretched O-H bond length through a process referred to as "bond softening".[47-51] The larger $R_{OH}$ leads to an increase in the dipole moment, μ, of water, which is given by $\vec{\mu} = \sum_i Q_i \vec{r}_i$ where $\vec{\mu}$ is the molecular dipole moment vector consisting of the charge $Q_i$ at the position $\vec{r}_i$ for the $i^{th}$ atom.[52] Thus, increasing the field strength of the laser field leads to a further stretching of the O-H bond length, which in turn increases the dipole moment of water and a stronger dipole interaction with the laser field. Field-dressing the molecule also reduces the bond angle which has a lower potential energy, leading to an increased dipole moment which points along the $C_2$ principal symmetry axis between the hydrogen atoms. Thus, the increase in the H-H internuclear distance, $R_{HH}$, is a consequence of the stretched O-H bond and slightly more bent H-O-H angle with increasing field strength. We calculate the dipole moment of water for the four FT-LIED measured $H_2O^+$ structures, the results of which are shown in Table 2.

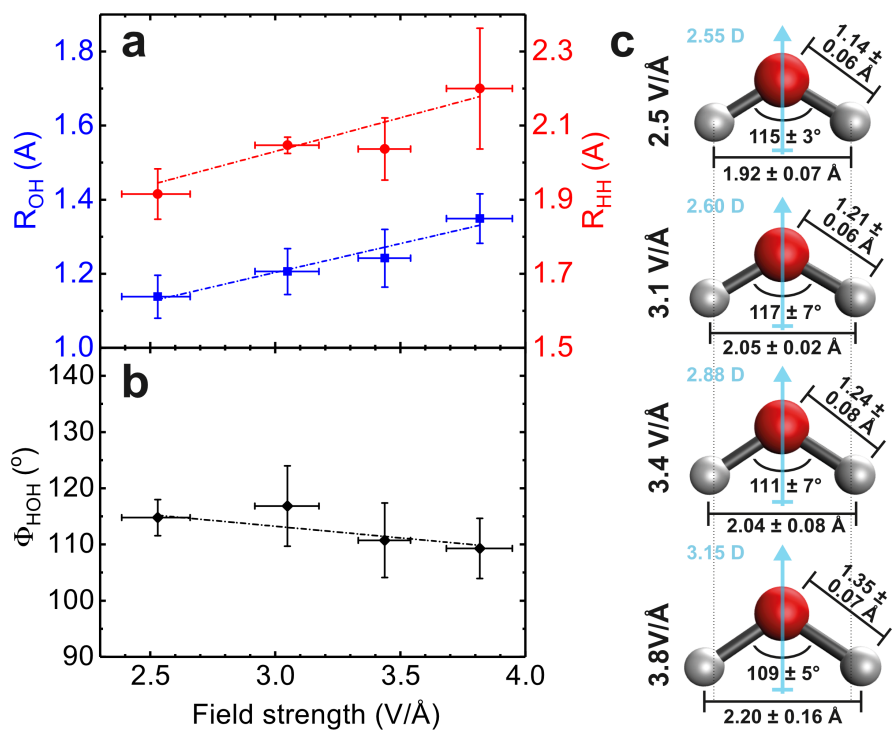

**Fig. 4 | Field-dependent stretching in $H_2O^+$.** (a) The O-H (blue) and H-H (red) internuclear distances, $R_{OH}$ and $R_{HH}$, respectively, and (b) H-O-H bond angle (black), $\Phi_{HOH}$, of $H_2O^+$ extracted with FT-LIED as a function of the laser field strength. A linear least-squares fit is applied (dash-dot lines). (c) The schematic of $H_2O^+$ with corresponding $R_{OH}$, $R_{HH}$ and $\Phi_{HOH}$ are shown at the various laser field strengths. The vertical dotted line is used to show the increase in the H-H internuclear distance with increasing field strength. The dipole moment of $H_2O^+$ is shown in light blue along with the corresponding calculated dipole moment values for the measured structures.

We calculate that upon increasing the field strength from 2.5 to 3.8 V/Å, an increase in the dipole moment of 8% to 33% is observed as compared to the equilibrium field-free $H_2O^+$ in its ground electronic state (2.37 D).[53,54] These increases in the calculated dipole moment of $H_2O^+$ coincide with an increase in the measured O-H internuclear distance of 14% and 35%. Shi *et al.* (2011) theoretically demonstrated that the molecule stretches and bends with increasing field strength when the molecule's dipole moment is aligned with the external field, further supporting our observed FT-LIED structural changes in water for field strengths of 2.5 – 3.8 V/Å. This is due to all electronic states (of the neutral and cation) being Stark-shifted in the presence of an external laser field, and leading to an altered ionization potential of water depending on whether the molecule's dipole moment is aligned parallel or anti-parallel to the electric field. In the case of parallel (anti-parallel) alignment, the ionization potential is lowered (increased) making it easier (more difficult) to ionize the water molecule.

**Table 2 | Calculated dipole moments for $H_2O^+$.** Dipole moments calculated for four geometries corresponding to the directly retrieved $H_2O^+$ structures shown in Fig. 4c.

| $R_{OH}$ (Å) | $\Phi_{HOH}$ (°) | Dipole moment (D) |
|---|---|---|
| 1.14 | 115 | 2.55 |
| 1.21 | 117 | 2.60 |
| 1.24 | 111 | 2.88 |
| 1.35 | 109 | 3.15 |

## IV – Summary and conclusions

In summary, we use FT-LIED to directly retrieve the geometric structure of $H_2O^+$ without *a priori* knowledge of the molecular structure nor the use of retrieval algorithms or *ab initio* calculations. We resolve a symmetrically stretched field-dressed $H_2O^+$ structure with picometre and femtosecond resolution using LIED that agrees best with a $H_2O^+$ structure in the ground electronic state. We use strong optical fields to additionally investigate the response of isolated water molecules to external fields at four field strengths between 2.5 – 3.8 V/Å. We observe that increasing the laser field intensity leads to an increase in the O-H bond length and a slight decrease in the H-O-H angle. These ultrafast structural changes lead to a larger dipole moment of water and a stronger coupling of the nuclear framework to the laser field.

Our results give insights into the nuclear response of an isolated water molecule that may illuminate the influence of strong optical fields on larger aggregates of water molecules, such as in the photochemistry of clusters and cells, where water plays an important role as a solvent or metabolite. In future work, the response of (water) clusters to external fields can be studied using LIED with the possibility of directly extracting inter- and intramolecular bond distances within a system of clusters.

**Acknowledgements**

We acknowledge financial support from the Spanish Ministry of Economy and Competitiveness (MINECO), through the "Severo Ochoa" Programme for Centres of Excellence in R&D (SEV-2015-0522) Fundació Cellex Barcelona and the CERCA Programme/Generalitat de Catalunya. X.L., K.A., T.S., A.S., B.B., M.S. and J.B. acknowledge the European Research Council (ERC) for ERC Advanced Grant TRANSFORMER (788218), MINECO for Plan Nacional FIS2017-89536-P, AGAUR for 2017 SGR1639, and Laserlab-Europe (EU-H2020 654148). K.A., J.B. and R. Moszynski acknowledge the Polish National Science Center within the project Symfonia, 2016/20/W/ST4/00314. X.L. and J.B. acknowledge financial support from China Scholarship Council. A.S. and J.B. acknowledge Marie Sklodowska-Curie Grant Agreement 641272. J. S. and S.G. acknowledge the ERC Consolidator Grant QUEMCHEM (772676). A.T.L. and C.D.L. are supported by the US Department of Energy under Grant DE-FG02-86ER13491.

**References**

[1] P. Ball, Chem. Rev. **108**, 74 (2008).
[2] L. Zhang, L. Wang, Y.-T. Kao, W. Qiu, Y. Yang, O. Okobiah, and D. Zhong, Proc. Natl. Acad. Sci. USA. **104**, 18461 (2007).
[3] S. Ebbinghaus, S. K. Kim, M. Heyden, X. Yu, U. Heugen, M. Gruebele, D. M. Leitner, and M. Havenith, Proc. Natl. Acad. Sci. USA. **104,** 20749 (2007).
[4] S. K. Sinha, S. Chakraborty, and S. Bandyopadhyay, J. Phys. Chem. B. **112**, 8203 (2008).
[5] S.-P. Shi, Q. Zhang, L. Zhang, R. Wang, Z.-H. Zhu, G. Jiang, and Y.-B. Fu, Chin. Phys. B **20**, 063102 (2011).
[6] D. H. Jung, J. H. Yang and M. S. Jhon, Chem. Phys. **244**, 331(1999).
[7] J. H. Sanderson, A. El-Zein, W. A. Bryan, W. R. Newell, A. J. Langley, and P. F. Taday, Phys. Rev. A **59**, R2567 (1999).

[8] F. Légaré, K. F. Lee, I. V. Litvinyuk, P. W. Dooley, S. S. Wesolowski, P. R. Bunker, P. Dombi, F. Krausz, A. D. Bandrauk, D. M. Villeneuve, and P. B. Corkum, Phys. Rev. A 71, 013415 (2005).

[9] M. Lein, J. P. Marangos, and P. L. Knight, Phys. Rev. A **66**, 051404 (2002).

[10] M. Meckel, D. Comtois, D. Zeidler, A. Staudte, D. Pavičić, H. C. Bandulet, H. Pépin, J. C. Kieffer, R. Dörner, D. M. Villeneuve, P. B. Corkum, Science **320**, 1478 (2008).

[11] C. D. Lin, A.-T. Le, Z. Chen, T. Morishita, and R. Lucchese, J. Phys. B: At., Mol. Opt. Phys. **43**, 122001 (2010).

[12] M. Okunishi, H. Niikura, R. R. Lucchese, T. Morishita, and K. Ueda, Phys. Rev. Lett. **106**, 063001 (2011)

[13] C. I. Blaga, J. Xu, A. D. DiChiara, E. Sistrunk, K. Zhang, P. Agostini, T. A. Miller, L. F. DiMauro, and C. D. Lin, Nature **483**, 194 (2012).

[14] J. Xu, C. I. Blaga, K. Zhang, Y. H. Lai, C. D. Lin, T. A. Miller, P. Agostini, L. F. DiMauro, Nat. Commun. **5**, 4635 (2014).

[15] M. G. Pullen, B. Wolter, A.-T. Le, M. Baudisch, M. Hemmer, A. Senftleben, C. D.Schröter, J. Ullrich, R. Moshammer, C. D. Lin, and J. Biegert, Nat. Commun. **6**, 7262 (2015).

[16] M. G. Pullen, B. Wolter, A.-T. Le, M. Baudisch, M. Sclafani, H. Pires, C. D. Schröter, J. Ullrich, R. Moshammer, T. Pfeifer, C. D. Lin, and J. Biegert, Nat. Commun. **7**, 11922 (2016).

[17] B. Wolter, M. G. Pullen, A.-T. Le, M. Baudisch, K. Doblhoff-Dier, A. Senftleben, M. Hemmer, C. D. Schröter, J. Ullrich, T. Pfeifer, R. Moshammer, S. Gräfe, O. Vendrell, C. D. Lin, J. Biegert, Science **354**, 308-312, (2016).

[18] Y. Ito, C. Wang, A.-T. Le, M. Okunishi, D. Ding, C. D. Lin, and K. Ueda, Struct. Dyn. **3**, 034303 (2016).

[19] K. Amini, M. Sclafani, T. Steinle, A.-T. Le, A. Sanchez, C. Müller, J. Steinmetzer, L. Yue, J. R. M. Saavedra, M. Hemmer, M. Lewenstein, R. Moshammer, T. Pfeifer, M. G. Pullen, J. Ullrich, B. Wolter, R. Moszynski, F. J. G. de Abajo, C. D. Lin, S. Gräfe, and J. Biegert, Proc. Natl. Acad. Sci. U.S.A. **116**, 8173 (2019).

[20] M. Dantus, S. B. Kom, J. C. Williamson, and A. H. Zewail, J. Phys. Chem. **98**, 2782 (1994).

[21] B. J. Siwick, J. R. Dwyer, R. E. Jordan, and R. J. D. Miller, Science **302** 1382 (2003).

[22] J. Yang, X. Zhu, T. J. A. Wolf, Z. Li, J. P. F. Nunes, R. Coffee, J. P. Cryan, M. Gühr, K. Hegazy, T. F. Heinz, K. Jobe, R. Li, X. Shen, T. Veccione, S. Weathersby, K. J. Wilkin, C. Yoneda, Q. Zheng, T. J. Martinez, M. Centurion, X. Wang, Science **361**, 64 (2018).

[23] T. J. A. Wolf, D. M. Sanchez, J. Yang, R. M. Parrish, J. P. F. Nunes, M. Centurion, R. Coffee, J. P. Cryan, M. Gühr, K. Hegazy, A. Kirrander, R. K. Li, J. Ruddock, X. Shen, T. Vecchione, S. P. Weathersby, P. M. Weber, K. Wilkin, H. Yong, Q. Zheng, X. J. Wang, M. P. Minitti, and T. J. Martínez, Nat. Chem. **11**, 504 (2019).

[24] B. Wolter, M. G. Pullen, M. Baudisch, M. Sclafani, M. Hemmer, A. Senftleben, C. D. Schröter, J. Ullrich, R. Moshammer, and J. Biegert, Phys. Rev. X **5**, 021034 (2015).

[25] M. Baudisch, B. Wolter, M. G. Pullen, M. Hemmer, and J. Biegert, Opt. Lett. **41**, 3583 (2016).

[26] U. Elu, M. Baudisch, H. Pires, F. Tani, M. H. Frosz, F. Köttig, A. Ermolov, P. St. J. Russell, and J. Biegert, Optica **4**, 1024 (2017).

[27] L. V. Keldysh, Sov. Phys. JETP **20** 1037 (1965).

[28] J. Ullrich, R. Moshammer, A. Dorn, R. Dörner, L. Ph. H. Schmidt, and H. Schmidt-Böcking, Rep. Prog. Phys. **66**,1463 (2003).

[29] H. J. A. Jensen, and H. Ågren, Chem. Phys. Lett. **110**, 140 (1984).

[30] H. J. A. Jensen, and H. Ågren, Chem. Phys. **104**, 229 (1986).

[31] H. J. A. Jensen and P. Jørgensen, and H. Ågren, J. Chem. Phys. **87**, 451 (1987).

[32] K. Aidas, C. Angeli, K. L. Bak, V. Bakken, R. Bast, L. Boman, O. Christiansen, R. Cimiraglia, S. Coriani, P. Dahle, E. K. Dalskov, U. Ekström, T. Enevoldsen, J. J. Eriksen, P. Ettenhuber, B. Fernández, L. Ferrighi, H. Fliegl, L. Frediani, K. Hald, A. Halkier, C. Hättig, H. Heiberg, T. Helgaker, A. C. Hennum, H. Hettema, E. Hjertenæs, S. Høst, I.-M. Høyvik, M. F. Iozzi, B. Jansík, H. J. A. Jensen, D. Jonsson, P. Jørgensen, J. Kauczor, S. Kirpekar, T. Kjærgaard, W. Klopper, S. Knecht, R. Kobayashi, H. Koch, J. Kongsted, A. Krapp, K. Kristensen, A. Ligabue, O. B. Lutnæs, J. I. Melo, K. V. Mikkelsen, R. H. Myhre, C. Neiss, C. B. Nielsen, P. Norman, J. Olsen, J. M. H. Olsen, A. Osted, M. J. Packer, F. Pawlowski, T. B. Pedersen, P. F. Provasi, S. Reine, Z. Rinkevicius, T. A. Ruden, K. Ruud, V. V. Rybkin, P. Sałek, C. C. M. Samson, A. S. de Merás, T. Saue, S. P. A. Sauer, B. Schimmelpfennig, K. Sneskov, A. H. Steindal, A. O. Sylvester-Hvid, P. R. Taylor, A. M. Teale, E. I. Tellgren, D. P. Tew, A. J. Thorvaldsen, L. Thøgersen, O. Vahtras, M. A. Watson, D. J. D. Wilson, M. Ziolkowski, and H. Ågren, Wiley Interdiscip. Rev. Comput. Mol. Sci. **4**, 269 (2014).

[33] J. Olsen, and P. Jørgensen, J. Chem. Phys. **82**, 3235 (1985).

[34] H. Fuest, Y. H. Lai, C. I. Blaga, K. Suzuki, J. Xu, P. Rupp, H. Li, P. Wnuk, P. Agostini, K. Yamazaki, M. Kanno, H. Kono, M. F. Kling, and L. F. DiMauro Phys. Rev. Lett. **122**, 053002 (2019).

[35] J. Kaiser, and R. Schafer, IEEE Trans. Acoust., Speech, Signal Process. **28**, 105 (1980).

[36] J. O. Smith III, "*Mathematics of the discrete Fourier transform (DFT): with audio applications*", Julius Smith (2007).

[37] J. B. Hasted, "Liquid water: Dielectric properties" The Physics and Physical Chemistry of Water. Springer, 255 (1972).

[38] A. G. Császára, G. Czakó, and T. Furtenbacher, J. Chem. Phys. **122**, 214305 (2005).

[39] T. R. Huet, C. J. Pursell, W. C. Ho, B. M. Dinelli, and T. Oka, J. Chem. Phys. **97**, 5977 (1992).

[40] K. Yuan, R. N. Dixon, and X. Yang, Acc. Chem. Res. **44**, 369 (2011).

[41] A. Baumann, S. Bazzi, D. Rompotis, O. Schepp, A. Azima, M. Wieland, D. Popova-Gorelova, O. Vendrell, R. Santra, and M. Drescher, Phys. Rev. A **96**, 013428 (2017).

[42] J. Suárez, L. Méndez, and I. Rabadán, Phys. Chem. Chem. Phys. **20**, 28511 (2018).

[43] C. G. Ning, B. Hajgató, Y. R. Huang, S. F. Zhang, K. Liu, Z. H. Luo, S. Knippenberg, J. K. Deng, and M. S. Deleuze, Chem. Phys. **343**, 19 (2008).

[44] S. Petretti, A. Saenz, A. Castro, and P. Decleva, Chem. Phys. **414**, 45 (2013).

[45] J. P. Farrell, S. Petretti, J. Förster, B. K. McFarland, L. S. Spector, Y. V. Vanne, P. Decleva, P. H. Bucksbaum, A. Saenz, and M. Gühr, Phys. Rev. Lett. **107**, 083001 (2011).
[46] B. J. Rao and A. J. C. Varandas, Phys. Chem. Chem. Phys. **17**, 6545 (2015).
[47] A. D. Bandrauk and M. L. Sink, J. Chem. Phys. **74**, 1110 (1981).
[48] P. H. Bucksbaum, A. Zavriyev, H. G. Muller, and D. W. Schumacher, Phys. Rev. Lett. **64**, 1883 (1990).
[49] A. Zavriyev, P. H. Bucksbaum, H. G. Muller, and D. W. Schumacher, Phys. Rev. A **42**, 5500 (1990).
[50] A. Zavriyev, P. H. Bucksbaum, J. Squier, and F. Salin, Phys Rev. Lett. **70**, 1077 (1993).
[51] H. Ibrahim, C. Lefebvre, A. D. Bandrauk, A. Staudte, and F. Légaré, J. Phys. B: At. Mol. Opt. Phys. **51**, 042002 (2018).
[52] P. W. Atkins, and R. S. Friedman, "*Molecular quantum mechanics*", Oxford University Press (2011).
[53] S. Wu, Y. Chen, X. Yang, Y. Guo, Y. Liu, Y. Li, R. J. Buenker, and P. Jensen, J. Mol. Spectrosc. **225**, 96 (2004).
[54] S. M. Nkambule, Å. Larson, S. F. dos Santos, and A. E. Orel, Phys. Rev. A **92**, 012708 (2015).